\documentclass[conference]{IEEEtran}
\IEEEoverridecommandlockouts
\usepackage{cite}
\usepackage{amsmath,amssymb,amsfonts}
\usepackage{algorithmic}
\usepackage{graphicx}
\usepackage{textcomp}
\usepackage{xcolor}
\def\BibTeX{{\rm B\kern-.05em{\sc i\kern-.025em b}\kern-.08em
    T\kern-.1667em\lower.7ex\hbox{E}\kern-.125emX}}
\begin{document}

\title{Improving Speaker Representations Using Contrastive Losses on Multi-scale Features}

\author{\IEEEauthorblockN{Satvik Dixit}
\IEEEauthorblockA{\textit{Dept. of Electrical and Computer Engineering} \\
\textit{Carnegie Mellon University}\\
Pittsburgh, USA \\
satvikdixit@cmu.edu}
\and
\IEEEauthorblockN{Massa Baali}
\IEEEauthorblockA{\textit{Language Technologies Institute} \\
\textit{Carnegie Mellon University}\\
Pittsburgh, USA \\
mbaalih@andrew.cmu.edu}
\and
\IEEEauthorblockN{Rita Singh}
\IEEEauthorblockA{\textit{Language Technologies Institute} \\
\textit{Carnegie Mellon University}\\
Pittsburgh, USA \\
rsingh@cs.cmu.edu}
\and
\IEEEauthorblockN{Bhiksha Raj}
\IEEEauthorblockA{\textit{Language Technologies Institute} \\
\textit{Carnegie Mellon University}\\
Pittsburgh, USA \\
bhiksha@cs.cmu.edu}
}

\maketitle

\begin{abstract}
Speaker verification systems have seen significant advancements with the introduction of Multi-scale Feature Aggregation (MFA) architectures, such as MFA-Conformer and ECAPA-TDNN. These models leverage information from various network depths by concatenating intermediate feature maps before the pooling and projection layers, demonstrating that even shallower feature maps encode valuable speaker-specific information. Building upon this foundation, we propose a Multi-scale Feature Contrastive (MFCon) loss that directly enhances the quality of these intermediate representations.
Our MFCon loss applies contrastive learning to all feature maps within the network, encouraging the model to learn more discriminative representations at the intermediate stage itself. By enforcing better feature map learning, we show that the resulting speaker embeddings exhibit increased discriminative power. Our method achieves a 9.05\% improvement in equal error rate (EER) compared to the standard MFA-Conformer on the VoxCeleb-1O test set. 
\end{abstract}


\section{Introduction}

Speaker representation learning is a critical component in speaker verification where the goal is to generate embeddings that can effectively distinguish between speakers. Ideally, embeddings belonging to the same speaker should cluster closely together, while those from different speakers should be well separated. Achieving this objective is crucial for developing robust speaker verification systems that can perform reliably across varying conditions and datasets.

Recent advances in speaker verification have seen the integration of contrastive learning techniques, which have significantly improved the discrimination power of speaker embeddings \cite{9413351, stafylakis2019selfsupervisedspeakerembeddings, xia2021selfsupervisedtextindependentspeakerverification, li2022speakerrepresentationlearning, li2023discriminativespeakerrepresentation, lepage2024additivemargincontrastiveselfsupervised, Lepage2023ExperimentingWA}. Contrastive learning encourages the model to learn representations where positive pairs (such as embeddings from same speakers and their augmentations) are pulled together, and negative pairs (such as embeddings from different speakers) are pushed apart. In \cite{li2022speakerrepresentationlearning}, the authors add an auxiliary contrastive loss term to the loss function for the ECAPA-TDNN \cite{Desplanques_2020} architecture and show an improvement in the speaker verification performance on the CN-Celeb dataset \cite{fan2019cncelebchallengingchinesespeaker}. 

Meanwhile, Multi-scale Feature Aggregation (MFA) has emerged as an effective technique for capturing speaker-specific information at different temporal resolutions. The ECAPA-TDNN architecture, for instance, demonstrated the benefits of multi-scale context aggregation in speaker verification tasks. Similarly, the ERes2Net \cite{chen2023enhancedres2net} framework showcased improved performance on short-duration speaker verification by employing multi-scale feature fusion. These approaches leverage the hierarchical nature of neural networks to extract and combine features at intermediate levels, resulting in more discriminative speaker representations.

In this paper, we propose a novel approach that combines the strengths of feature aggregation and contrastive learning. We introduce a Multi-Scale Feature Contrastive (MFCon) loss that applies contrastive learning across all the intermediate feature maps. The training objective of our model is a combination of Additive Margin Softmax (AM-Softmax) loss  \cite{Wang_2018} on the speaker embeddings and a Supervised Contrastive Learning (SupCon) loss  \cite{khosla2021supervisedcontrastive} on the intermediate feature maps. This joint optimization strategy allows our model to balance the benefits of traditional classification-based learning with the discriminative power of contrastive learning. 

Experiments conducted on VoxCeleb1 \cite{Nagrani_2017} demonstrate the superiority of the proposed model over the previous ones. In addition, we conducted an ablation study to determine the contributions of different components of the proposed loss function. We looked at different coefficients for the contrastive loss term, different contrastive losses and different ways of organizing the architecture. Finally we combine the MFCon loss with a SupCon loss on the speaker embedding, this results in a substantial improvement of 9.05\% in EER on Voxceleb1-O evaluation benchmark. Code has been made publicly available at https://github.com/satvik-dixit/MFCon.

\section{Methodology}

\begin{figure*}[htbp]
\centering
\includegraphics[width=\linewidth]{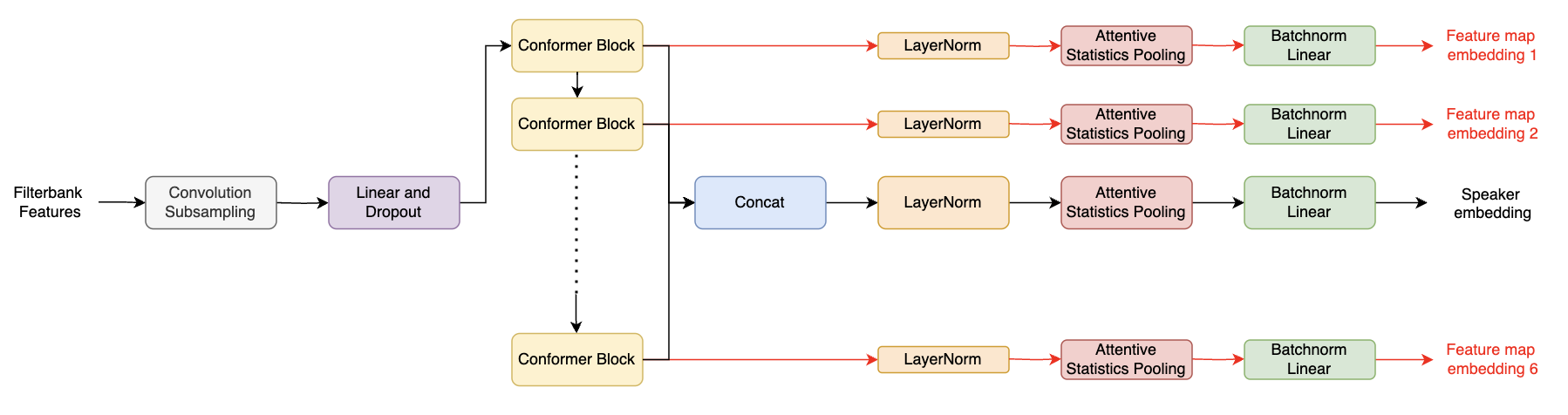}
\caption{The MFCon loss applied to the MFA conformer model. The arrows in black show the original MFA conformer and the arrows in red show the architecture to obtain embeddings from the feature maps of the Conformer blocks}
\label{fig:architecture}
\end{figure*}

Our proposed approach, which we call MFCon Loss, builds upon the MFA idea and incorporates supervised contrastive learning on the intermediate feature maps to enhance speaker representations. Our hypothesis is that explicitly enforcing the separation of intermediate speaker representations at each scale will also enhance the overall speaker embedding. This section details the components of our method, including the architecture and the novel loss function.

\subsection{Architecture}

We implement our proposed loss function on the MFA-Conformer \cite{zhang2022mfaconformer} architecture, a state-of-the-art framework for speaker verification tasks. The MFA-Conformer comprises of a series of Conformer \cite{gulati2020conformerconvolutionaugmentedtransformerspeech} blocks, each capturing distinct levels of acoustic information. Our key contribution lies in the way we use these output feature maps to enhance the final speaker representations.

To apply contrastive loss to the conformer block outputs, we derive embeddings from each block using a process similar to the original MFA-Conformer. For each Conformer block, we extract the output feature map and apply layer normalization. These normalized feature maps then undergo attentive statistics pooling \cite{Okabe_2018}, which computes weighted means and standard deviations across the temporal dimension, effectively aggregating information into fixed-dimensional vectors. Finally, we employ batch normalization followed by a linear projection to obtain the final feature map embedding for each Conformer block. 

\subsection{Loss Function Components}

Our MFCon Loss combines two key components: AM-Softmax loss and Supervised Contrastive Loss.

\subsubsection{AM-Softmax Loss}

The AM-Softmax loss is applied to the final speaker embedding. This is the same loss being used in the original MFA Conformer. This loss function enhances the discriminative power of the learned representations by introducing an additive margin in the angular space. The AM-Softmax loss is defined as:

\begin{equation}
L_{AMS} = -\frac{1}{N} \sum_{i=1}^N \log \frac{e^{s(\cos(\theta_{y_i} + m))}}{e^{s(\cos(\theta_{y_i} + m))} + \sum_{j \neq y_i} e^{s\cos(\theta_j)}}
\end{equation}

where $N$ is the batch size, $s$ is the scale factor, $m$ is the additive margin, $\theta_{y_i}$ is the angle between the feature vector and the weight vector of the ground truth class, and $\theta_j$ represents the angles with other class weight vectors.

\subsubsection{Supervised Contrastive Loss}

We apply the Supervised Contrastive (SupCon) loss to the feature map embeddings. This loss encourages the model to learn representations where samples from the same speaker are pulled closer together in the embedding space, while samples from different speakers are pushed apart. The loss is defined as:

\begin{equation}
L_{SupCon} = \sum_{i \in I} \frac{-1}{|P(i)|} \sum_{p \in P(i)} \log \frac{\exp(z_i \cdot z_p / \tau)}{\sum_{a \in A(i)} \exp(z_i \cdot z_a / \tau)}
\end{equation}

where $I$ is the set of indices, $P(i)$ is the set of positives distinct from $i$, $A(i)$ is the set of all indices distinct from $i$, $z_i$ are the normalized embeddings, and $\tau$ is a temperature parameter.

\subsubsection{Multi-scale Feature Contrastive (MFCon) Loss}

Our final loss function, MFCon Loss, combines the AM-Softmax loss applied to the speaker embedding and the mean of the Supervised Contrastive losses applied to the feature maps:

\begin{equation}
L_{MFCon} = L_{AMS} + \lambda \frac{1}{L} \sum_{i=1}^{L} L_{SupCon}^{(i)}
\end{equation}

where $\lambda$ is a hyperparameter that balances the contribution of the contrastive loss component, L is the number of conformer blocks and $L_{SupCon}^{(i)}$ is the supervised contrastive loss applied to the $i-th$ feature map.

\begin{figure}[htbp]
    \centering
    \includegraphics[width=\linewidth]{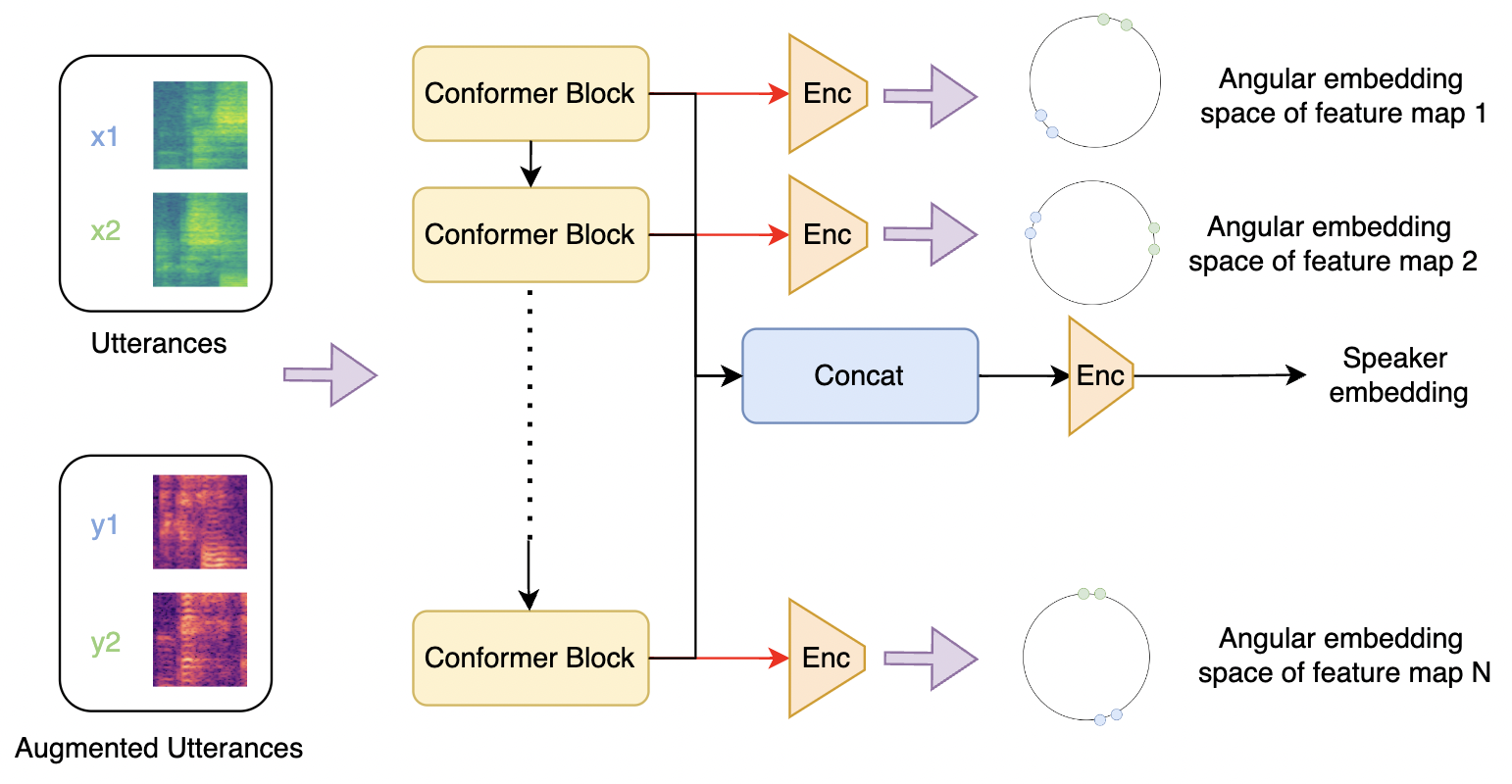} 
    \caption{
    Illustration of the MFCon idea. 
    Utterances x1 and x2 from distinct speakers, along with their augmentations y1 and y2, are processed through N conformer blocks. Feature map embeddings are extracted at each scale using dedicated encoders. The MFCon loss applies contrastive learning across all scales, simultaneously minimizing intra-speaker distances and maximizing inter-speaker distances in the embedding space}
    \label{fig:idea} 
\end{figure}

The introduced contrastive loss enforces a crucial constraint on the feature map embeddings: those derived from the same speaker and their augmentations are pulled closer together in the embedding space, while embeddings from different speakers are pushed apart as shown in \ref{fig:idea}. This mechanism ensures that the Conformer blocks learn to generate more discriminative and speaker-specific feature maps at multiple scales.

Our proposed approach combines AM-Softmax loss applied to the final speaker embedding with supervised contrastive loss applied to the feature map embeddings. This dual-objective optimization strategy leverages the complementary strengths of both loss functions. The AM-Softmax loss provides strong supervision for the final embedding, encouraging clear decision boundaries between speaker classes. Simultaneously, the contrastive loss enhances the quality of intermediate representations across different scales. 

\section{Experiments and Results}

\subsection{Dataset and Evaluation Metrics}

We use VoxCeleb1, a large-scale publicly available audio-visual dataset of human speech extracted from YouTube interview videos. For model training, we utilize the development set of VoxCeleb1, which contains over 148,000 utterances from 1,211 speakers. 

For evaluation, we employ the Vox1-O protocol, which is the most widely used evaluation benchmark in recent speaker recognition research. This protocol utilizes approximately 37,000 trials from 40 speakers in the VoxCeleb1 test set. 

    

The performance was measured using Equal Error Rate (EER) and minimum Detection Cost Function (minDCF) with $p_{target} = 0.01$.

\subsection{Implementation Details}

We implement our proposed architecure and loss function using the PyTorch framework. The model architecture and training procedure is the same as described in the original MFA Conformer paper \cite{zhang2022mfaconformer}. The model architecture includes 6 Conformer blocks with subsampling rates  equal to 1/2. We use 80-dimensional Fbanks as input features. During training, we also take randomly extracted 3-second segments from each utterance in the dataset. We trained the model using the Adam optimizer, starting with an initial learning rate of 0.001, which is reduced by 50\% every 5 epochs. The batchsize is fixed as 100.

The AM-Softmax loss function component has a margin of 0.2 and a scaling factor of 30. The SupCon loss has temperature of 0.07.
For the feature map embedding extraction, we maintain the same architectural components as in the original MFA-Conformer, including layer normalization, attentive statistics pooling, batch normalization, and a linear projection. However, we adjust the embedding dimension to 192 to accommodate the feature maps from individual Conformer blocks, as opposed to the 192 x 6 dimensionality used for the concatenated feature maps in the original design.

During training, we apply augmentations. For each input signal in the batch, we create an augmented version and add it to the batch. The augmentation process randomly applies either noise or reverberation to the utterance. For noise augmentation, we use the MUSAN corpus \cite{snyder2015musan}, which provides a diverse set of background noises, overlapping music tracks, and speech segments. For reverberation, we use the simulated Room Impulse Response database \cite{Ko2017}.  


\subsection{Comparison of Contrastive Loss Functions}

To assess the impact of different contrastive loss functions on our method, we experimented with coomonly used contrastive loss functions while keeping the rest of the architecture constant. The coefficient is fixed as 0.1. Table \ref{tab:contrastive_loss_comparison} presents the results on the VoxCeleb1-O benchmark for four widely-used contrastive loss functions.

\begin{table}[ht]
\centering
\caption{Comparison of contrastive loss functions on VoxCeleb1-O}
\label{tab:contrastive_loss_comparison}
\begin{tabular}{lcc}
\hline
Contrastive Loss & \multicolumn{2}{c}{VoxCeleb1-O} \\
 & EER (\%) & minDCF \\
\hline
Triplet Loss & 2.61 & \textbf{0.27} \\
N-pair Loss & 2.63 & 0.27 \\
NTXent Loss & 2.72 & 0.27\\
SupCon Loss & \textbf{2.56} & 0.28 \\
\hline
\end{tabular}
\end{table}

The results in Table \ref{tab:contrastive_loss_comparison} show that Supervised Contrastive (SupCon) loss achieves the best performance. 

\subsection{Ablation Studies}

To explore the importance of key components in our MFCon loss, we conducted some ablation studies. Specifically, we investigated the impact of parameter sharing in the pooling and projection layers across different feature maps. Table \ref{tab:ablation_studies} presents the results of these ablation experiments on the VoxCeleb1-O dataset

\begin{table}[ht]
\centering
\caption{Ablation studies on MFCon components on VoxCeleb1-O}
\label{tab:ablation_studies}
\begin{tabular}{lcc}
\hline
Configuration & \multicolumn{2}{c}{VoxCeleb1-O}  \\
 & EER (\%) & minDCF \\
\hline
Full MFCon Loss & \textbf{2.56} & 0.28 \\
Shared Pooling Layer& 2.72 & 0.28 \\
Shared Projection Layer & 2.65 & 0.30 \\
Shared Pooling and Projection Layers  & 2.60 & \textbf{0.27} \\
\hline
\end{tabular}
\end{table}

The ablation results demonstrate that having separate parameters for the pooling layer and projection layer results in the best performance.  

\subsection{Tuning the coefficient {$\lambda$}}

We experimented with 4 different coefficients for the contrastive loss term. Table \ref{tab:coeff_comparison} shows the results on the VoxCeleb1-O benchmark.

\begin{table}[ht]
\centering
\caption{Impact of contrastive loss coefficient ($\lambda$) on MFCon performance}
\label{tab:coeff_comparison}
\begin{tabular}{lcc}
\hline
Coefficient & \multicolumn{2}{c}{VoxCeleb1-O} \\
 & EER (\%) & minDCF \\
\hline
$\lambda$ = 0.01 & \textbf{2.52} & \textbf{0.27} \\
$\lambda$ = 0.03 & 2.61 & 0.27  \\
$\lambda$ = 0.1 & 2.56 & 0.28  \\
$\lambda$ = 0.3 & 2.71 & 0.29 \\
\hline
\end{tabular}
\end{table}

The results show that $\lambda$ = 0.01 achieves the best performance.

\subsection{Comparison with Baseline Methods}

Table \ref{tab:baseline_comparison} presents the evaluation results comparing MFCon with previous methods, including AM-Softmax and AM-SupCon.

\begin{table}[ht]
\centering
\caption{Comparison of MFCon with previous methods on VoxCeleb1-O }
\label{tab:baseline_comparison}
\begin{tabular}{lcc}
\hline
Method & \multicolumn{2}{c}{VoxCeleb1-O} \\
 & EER (\%) & minDCF \\
\hline
AM-Softmax & 2.65 & 0.29  \\
AM-SupCon & 2.56 & 0.26  \\
MFCon & \textbf{2.52} & \textbf{0.26} \\
\hline
\end{tabular}
\end{table}

The AM-SupCon method \cite{Wang_2018}, originally proposed by Li et al. for the ECAPA-TDNN architecture, combines a Supervised Contrastive (SupCon) Loss on the speaker embedding with the standard AAM-Softmax loss \cite{Deng_2022}. We adapted this approach for the MFA-Conformer architecture, referring to it as AM-SupCon in our experiments. . We see an improvement over standard AMSoftmax Loss. MFCon outperforms both of these models with an EER of 2.52\%. 

Both AM-SupCon and MFCon losses include a coefficient ($\lambda$) for the contrastive loss term. The reported results were obtained after tuning this hyperparameter over the values {0.01, 0.03, 0.1, 0.3} to ensure optimal performance for each method. As shown in Table \ref{tab:baseline_comparison}, MFCon outperforms previous approaches, demonstrating the effectiveness of improving intermediate feature maps on learning discriminative speaker representations.

\subsection{Combining AMSupcon and MFCon}

We combined the AMSupcon loss and MFCon loss to see the effect of using contrastive learning on both the speaker embedding and the intermediate feature maps. The combined loss is:
\begin{equation}
Loss = L_{AMSoftmax} + \lambda_{1}\frac{1}{L} \sum_{i=1}^{L} L_{SupCon_1}^{(i)} + \lambda_{2} L_{SupCon_2}
\end{equation}
Here $L_{SupCon_1}$ is applied to the feature maps and $L_{SupCon_2}$ is applied to the speaker embedding.
 We conducted a series of experiments to optimize the coefficients $\lambda_{1}$ and $\lambda_{2}$ and compare the performance with individual MFCon and AMSupCon approaches. Table \ref{tab:combined_ablation_studies} presents the results of these experiments.

\begin{table}[ht]
\centering
\caption{Ablation studies to see the effect of combining MFCon and AMSupCon on VoxCeleb1-O}
\label{tab:combined_ablation_studies}
\begin{tabular}{lcccc}
\hline
Configuration & $\lambda_{1}$ & $\lambda_{2}$ & \multicolumn{2}{c}{VoxCeleb1-O}  \\
& & & EER (\%) & minDCF \\
\hline
MFCon Loss & 0.01 & 0.00 & 2.52 & 0.27 \\
AMSupCon Loss & 0.00 & 0.03 & 2.56 & 0.26 \\
Combined Loss & 0.01 & 0.01 & 2.47 & 0.27 \\
Combined Loss & 0.03 & 0.03 & \textbf{2.41} & \textbf{0.25}\\
Combined Loss  & 0.10 & 0.10 & 2.63 & 0.27 \\
Combined Loss  & 0.30 & 0.30 & 2.72 & 0.28 \\
\hline
\end{tabular}
\end{table}

The ablation results in Table \ref{tab:combined_ablation_studies} show that for $\lambda_{1}$ = 0.03 and $\lambda_{2}$ = 0.03, we get an EER of 2.41\% which is a 9.05\% improvement over the AMSoftmax baseline EER of 2.65\%. The performance of the combined loss is significantly better than using MFCon alone or AMSupcon alone where the EERs are 2.52\% and 2.56\% respectively. This effect can be explained by complimentary nature of both the losses where MFCon improves the quality of the intermediate feature maps and  AMSupCon enhances the discriminative ability of the speaker embedding.

\section{Conclusion}

In this paper, we introduced the Multi-scale Feature Contrastive (MFCon) loss, a novel approach that significantly enhances speaker verification systems by leveraging contrastive learning on multi-scale features. We find that explicitly enhancing the speaker separability of the intermediate feature maps improves the discriminative ability of the final speaker embedding. The proposed MFCon loss demonstrated substantial improvements in speaker verification performance, achieving a 4.91\% reduction in equal error rate on the VoxCeleb-1O evaluation benchmark compared to the standard MFA-Conformer. Furthermore, we showed that combining MFCon with AMSupCon loss yields even greater performance improvements of 9.05\%, highlighting the complementary nature of contrastive learning at different levels of the network. Through comprehensive ablation studies, we identified optimal configurations for the contrastive loss. Our work underscores the significance of optimizing intermediate representations in  speaker verification tasks and provides a flexible framework for future research.

\newpage

\bibliographystyle{IEEEtran}
{\large\bibliography{refs}}

\end{document}